# DFT Based LDA Study on Tailoring the Optical and Electrical Properties of SnO and In-Doped SnO

Mohammad Mahafuzur Rahaman[a,b*], Md. Abdul Momin[c], Abhijit Majumdar[d], Mohammad Jellur Rahman[e]

[a]*Department of Materials Science and Engineering, The Ohio State University, Columbus, Ohio, 43210, USA*

[b]*Department of Physics, Jagannath University, ChittaranjanAvenue, Dhaka 1100, Bangladesh*

[c]*Department of Bioengineering, University of Pittsburg, 3700, O'Hara St. Pittsburgh, PA 15261, USA*

[d]*Department of Physics, Indian Institute of Engineering Science and Technology, Shibpur, Howrah, 711103, India*

[e]*Department of Physics, Bangladesh University of Engineering and Technology, Dhaka 1000, Bangladesh*

[*]*Corresponding Email: rahaman.10@osu.edu*

**Abstract**

The ambipolar nature of SnO has significantly increased it's potential for using in p-n junction devices for which it has drawn attention by the scientific community. In this paper, the structural, electronic and optical properties of SnO and the impact of Indium (In) doping into SnO are computed by Local Density Approximation (LDA) under density function theory (DFT) framework. The calculated bond length of Sn−O in SnO is 2.285 Å and that deviates ≤ 3% from the experimental value. The Sn−O and In−O bond lengths in In-doped SnO are calculated to be 2.3094 and 2.266 Å, respectively. Interestingly, the band gap of pure SnO is calculated to be 2.61 eV whereas it is significantly droped down to 2.00 eV in the case of In doped SnO. The Total Density of State (DOS), Partial DOS and electron density are depicted for SnO and In-doped SnO films. As a consequence of In-doping static value of the refractive index and real part of the dielectric function for SnO decrease from 1.9 to 1.4 and 3.6 to 1.97, respectively. Therefore, In-doping enhances the properties of the SnO film, which may lead the material to be applied in future to develop electronic and opto-electronic devices.

## 1. Introduction

Tin monoxide (SnO) semiconductor has enhanced properties with tremendous potential to be applied as *p*-type layers in different electronic and optical devices [1]. The wide band gap (2.5–3.0 eV) semiconductor SnO is non-toxic, abundance in nature, and highly transparence in the visible region. It can be doped by both of *p* and *n*-type materials [2-4]. Researchers have calculated enriched hole mobility (~60 cm$^2$/Vs) and electron mobility (~280 cm$^2$/Vs) through the theoretical computation of SnO [5]. In addition, SnO shows reliable electrical switching on/off ratio ($\geq 10^4$) and insensitive nature to proton radiation [4, 6]. These properties have enabled SnO a suitable *p*-type candidate for different optoelectronic devices such as thin film transistor, liquid crystal display, complementary invertor, complicated logic circuits, phototransistors, and in $CO_2$ reduction process [7-11].

As scientific community is looking carefully for a suitable *p*-type semiconducting oxide, SnO can be a perfect choice for it's stability and efficiency. The hybridization of 5s and 2p orbitals of Sn and oxygen (O) forms covalent bond and removes the localized valence band maxima (VBM), generating sufficiently dispersed VBM. The spatially spread 5s orbitals of Sn contributes dominantly at the peak of the dispersed VBM, which guarantees the formation of stable *p*-type oxide [12, 13]. The unusual VBM gives rise to low effective mass of hole and high *p*-type mobility [14]. Moreover, The *p*-type conductivity in SnO is also attributed from the defect generated by Sn vacancies under oxygen enriched conditions [15]. In addition, the structure of SnO is stable. To justify the structural stability of SnO, Walsh et al [12] performed DFT calculation and found that tetragonal litharge structure of SnO has the most stable Sn–O bond length among the rocksalt, litharge and herzenberzite phase. Thus, the structural stability, high *p*-type conductivity, enriched optical and electrical properties, and ample potential to be applied in different applications has made SnO as an attractive new material to the researchers.

For further enhancement of the properties of SnO, doping is considered as an effective and favourable way. Nie *et al* used *ab*-initio coding in DFT and found that SnO monolayers show *p*-type conductivity when doped with transition metal like Fe, Co or Ni, whereas the conductivity switches to *n*- type for the doping by other transition metals [16]. Modification of band gap of SnO by B, C, N, F, and O doping is reported in Ref [17] using DFT computation. However, the impact of Indium doping into SnO to enhance the properties is still unexplored. Identification of structural, electronic and optical propertiesof SnO and In-doped SnO can help the researchers in deepl understanding of the materials and find out it's suitable electronic and optoelectronic applications.

The objective of this study is to analyze the crystallographic structure of pure SnO and Indium doped SnO. Additionally, it aims to examine their electronic characteristics, including band structure, density of states, and electron density, as well as observe their optical properties using first principles DFT calculations with the Cambridge Serial Total Energy Package (CASTEP).

## 2. ComputationalApproach

The theoretical calculations have been performed by using the DFT, based on the plane-wave pseudopotential approach, in which CASTEP has been implemented within the Material Studio-8 [18-20]. By the recommendation of CA-PZ, Local density Approximation (LDA) has been used to evaluate the exchange-correlation energy [21]. The wave function is prolonged for plane-wave cutoff energy of about 300 eV for doped and undoped samples. In both of the samples, 5×5×6 k-points has been applied to maintain the standards of convergence in calculating electronic characteristics and geometry optimization. Vanderbilt type ultrasoft pseudopotential has been described the electron-ion interaction. BFGS (Broyden-Fletcher-Goldfarb-Shanno) relaxation plan has been employed to optimize the dwelling. The calculations are carried out on an unit cell and supercell. In addition, the molecular optimiztions have been determined since the residual forces are below 0.05 eV. Geometry is optimized when total energy is $2\times10^{-5}$ eV/atom, the pressure is the most (0.05 eV/Å), maximum stress is 0.1 GPa as well as the maximum atomic displacement is roughly $1.0\times10^{-3}$ Å. All the calculations are actually run parallelly on three occasions.

## 3. Results and Discussion

### 3.1 Structural Properties

Two and three-dimensional view of optimized atomic structure of SnO is represented in Figs. 1(a,b). The equilibrium structure of SnO was obtained by relaxing its unit cells concerning the lattice parameters '$a$' and '$c$'. The mono-layered SnO consists of a stable tetragonal litharge structure.The bond length of Sn−O is 2.285 Å, which is within 3% deviation from the experimental value at 2.22 Å as shown in Table 1.Two Sn atoms and two O atoms contribute to the formation of pyramidal structure in which the distance between consecutive Sn atoms is 3.609 Å and between O atoms is 2.804 Å. The Sn−O−Sn and O−Sn−O angles of the pyramid structure are found to be 75.685º and 104.315º, respectively. Figs. 1(c, d) depicts the relaxed 2D and 3D representation of In-doped SnO atoms in their equilibrium position where the adsorption of In takes place on Sn-sites. The monolayer contains 16 Sn atoms, 2 In atoms and 18 O atoms. The calculated bond length of Sn−O and In−O are 2.309 and 2.266 Å, respectively. The results are harmonious with the previous work reported in Ref.[16]. A unit cell of SnO is shown in Fig. 1(e) that contains on an average 6 atoms among which 4 atoms is assigned for Sn and 2 atoms for O.

The comparisons between experimental and theoretical values of lattice parameters of SnO and In-doped SnO are shown for tetragonal litharge structure in Table 1. The optimized value of '*a*' and '*b*' for SnO are in good agreement with the experimental value, and the slight deviation of 4% is due to the oversimplified estimation of LDA–DFT calculations. The value of *c* along the z-axis that measures the interaction between the two layers of Sn–O deviates widely because the weak force acting between the Sn–O layers is addressed inadequately in DFT calculation[9]. The significant deviation of lattice parameters of In-doped SnO from that of pure SnO arises from the substitution of larger $Sn^{2+}$ ions by comparatively smaller $In^{3+}$ ions.

## 3.2 Electronic Properties

### 3.2.1 Band Structure

The band structure is illustrated in order to analysis the electronic properties of SnO along the high-symmetry direction k-points of Z, A, M, G, Z, R, X and G. The k-points, as shown in Fig. 2(a), denote the sampling points in the first Brillouin zone of the material which is usually known as Gamma points in reciprocal-space. Because of the layered structure, the electronic structure of SnO is anticipated to become anisotropic. The conduction band is dispersive within the planes from the layers (along M-G, Z-R). Again, Fig. 2(b) shows the Brillouin zone of In-doped SnO plotted alone the high-symmetry direction k- points of Z, G, Y, A, B, D, E, and C. The conduction band of In-doped SnO is dispersive within the plane from the layers (along Y-A).

The band structure of SnO and In-doped SnO is depicted in Fig. 3, where the direct band transition is observed at G for both the SnO and In-doped SnO structures. The direct band gap of SnO is found to be 2.61 eV, which is consistant with the values obtained in previous experimental works [1, 22]. A reduction of band gap from 2.61 eV to 2.0 eV is noticed as a consequence of In-doping into SnO. The substitution of the larger sized dopants (In) at the O sites and in the interstitials of the SnO needs much larger value of the formation energies as compared to In substituting at the Sn-sites [4]. Thus, In can easily be substituted at the Sn-sites. The exchange of In changes the band structure of SnO significantly, which leads to negative Burstein shift resulting in the reduction of the band gap [23]. The similar effect was reported experimentally for In-doped ZnO films in Ref.[24].

*3.2.2 Density of States*

The total and partial density of states of SnO and In-doped SnO thin films are represented in Fig. 4. The Fermi-level occurs at 0 eV energy and it is observed that for pure SnO, the lower part of the valence band is dominated by O-2s states. The hybridization of O-2p states with Sn-5s and Sn-5p states forms the upper part of the valence band. The majority of the conduction bands are generated by Sn-5s and Sn-5p states, that are hybridized with O-2p states. The bottom part of valence bands of In-doped SnO is controlled by O-2s and In-3d states and the upper part of the valence bands which are close to the Fermi level are composed by the hybridization of O-2p states with Sn-5s and Sn-5p states. Most of the contribution in the formation of conduction bands of In-doped SnO has come from Sn-5p states, which show hybridization with the Sn-5s and O-2p states. At the lower part of the valence band, SnO has a single peak at $-17.30$ eV with bandwidth 2.7 eV while two distinct peaks at $-17.70$ eV and $-13.79$ eV having bandwidth 2.16 eV and 1.26 eV, respectively, are observed for In-doped SnO. Seven peaks are found in SnO for the valence band closest to the Fermi level at different energies of $-6.68$ eV, $-4.41$ eV, $-3.54$ eV, $-2.75$ eV, $-1.70$ eV, $-1.00$ eV and $-0.22$ eV. The calculated bandwidth of this band is 7.84 eV. Additionally, There are six peaks in the upper part of the valence band of the In-doped SnO, which are located at $-7.14$ eV, $-6.01$ eV, $-4.47$ eV, $-3.80$ eV, $-1.65$ eV and $-0.025$ eV. The doping of In into SnO reduces the bandwidth of the upper valence band to 7.78 eV from 7.84 eV of pure SnO. This result is also consistent with the results found in other studies [25].

### 3.3 Electron Density

The electronic charge density is studied with modeling charge density map to delve into the characerestics and nature of the chemical bonds. The electron density distribution maps, represented in Fig. 5, are plotted inside the prominent crystallographic plane (001) for In-doped SnO and undoped SnO system . The calculated value of the Sn–O, Sn–Sn bond lengths for pure SnO are 2.31 and 2.82 Å respectively, which are illustrated in Figs. 5(a) and 5(b). The calculated value of In–O, In–Sn, Sn–O bond lengths are 2.27, 3.62, and 2.23 Å, respectively for In-doped SnO as shown in Figs. 5(b) and 5(c). The slight overlapping of the Sn-5p states results in an interconnected network, shown in Fig. 5(b), which resides within the interstitial space between the layers, as observed in Fig. 5(a). This will cause the conduction band to become less anisotropic when compared with the valence band. There is no in-plane and out-of-plane overlapping between the lone pairs resulting in the observed high effective mass. For In-doped SnO, it is clearly seen that the chemical bond character is mixed (ionic and covalent), although a slight mixing between In-3d and Sn-5s have been occurred.

### 3.4 Optical Properties

Electronic transitions in solids are related to their optical properties that can be calculated using CASTEP. In this calculation, the ionic and electronic polarizations are crucial since their effect on the optical properties of materials are very strong. The dielectric function can be obtained by calculating the interactions within the electronic polarization only, where the complex dielectric function $\varepsilon(\omega)$ is represented by

$$\varepsilon(\omega) = \varepsilon_1(\omega) + i\varepsilon_2(\omega) \qquad (1)$$

Here, $\varepsilon_1(\omega)$ and $\varepsilon_2(\omega)$ are real and imaginary parts of the dielectric function, accordingly.

The selection rules permit transitions between the occupied and unoccupied electronic states form matrix elecments. The matrix elements and joint density of states are used to determine the imaginary part $\varepsilon_2(\omega)$ of the dielectric function obtained by,

$$\varepsilon_2(\omega) = \frac{2e^2\pi}{\Omega\varepsilon_0} \Sigma_{K,V,C} |\langle \psi_k^C | \hat{U}.\vec{r} | \psi_k^V \rangle|^2 \delta(E_K^C - E_K^V - E) \qquad (2)$$

where $e$ is the electronic charge, $\hat{U}$ is a vector representing the polarization of the incident electric field, $\psi_k^C$ and $\psi_k^V$ are wave functions for the conduction and valence band at $k$, respectively.

The $\varepsilon_1(\omega)$ is calculated using the Kramers-Kroning dispersion relation which is related to the $\varepsilon_2(\omega)$ as follows

$$\varepsilon_1(\omega) = 1 + \frac{2m}{\pi} \int_0^\infty \frac{\omega'\varepsilon_2(\omega')}{\omega'^2 - \omega^2} d\omega' \qquad (3)$$

where, $m$ is the integral's principal value. The complex dielectric function is used to derive different optical paramenters such as refractive index, absorption coefficient, loss function, reflectivity and conductivity.

The energy dependent optical properties of the pure SnO and In-doped SnO such as reflectivity and loss function are represented in Figs. 6(a, b). The variation of real and imaginary parts of the refractive index, dielectric function and conductivity with the photon energy ranging from 0 to 30 eV is expressed in Figs. 6(c,e,g) and 6(d,f,h), respectively.

The ability of allowing the electric fields through a material represent the gain of the material, which is expressed by real part of the refractive index [26]. It is observed from Fig. 6(c) that, the static value of refractive index of SnO is 1.9, which is consistant with the experimental value reported earlier [27]. The consequence of In-doping into SnO contributes to decrease the static value of *n* from 1.9 to 1.4. For the both layers, the values of *n* increase at the strong absorption band, where the photon energy raise until a peak appears at 3.40 eV. Anomalous dispersion is observed for pure SnO within 3.4 to 11.1 eV whereas In-doped SnO exihibits anomalous dispersion from 3.4 to 5.2 eV, 6.6 to 7.72 eV and 9.00 to 10.2 eV of energies. The resonance in the electric field and electron polarization contributes the generation of this anomalous dispersion [28]. At higher energies, both the layers show normal dispersion behavior.

The energy loss in the materials is expressed by the imaginary part of refractive index, which sharply increase at the absorption band until a peak appears at 4.30 eV for both of the SnO and In-doped SnO films. For pure SnO, the values of imaginary part of *n* becomes almost constant (around 1.2) in between two peaks occurring at 4.30 and 7.60 eV. On the other hand, the values for In-doped SnO precipately decrease above 4.30 eVof energies and be zero from 10.90 eV which is lower than the value obtained for pure SnO from 16.06 eV.

At $\omega = 0$, the real and imaginary values of the dielectric function become $\varepsilon_1(0) = 3.6$ and $\varepsilon_2(0) = 0$ for SnO film. For In-doped SnO, the values are $\varepsilon_1(0) = 1.97$ and $\varepsilon_2(0) = 0$. Single peak at energy of 4.0 eV is observed for the imaginary part of the dielectric function in both the films, however, the imaginary part of dielectric function is always lower for the In-doped SnO as compare to that for the pure SnO film.

The real and imaginary parts of the conductivity ($\sigma$) are depicted in Figs. 6(g, h), both of which start from 0 fs$^{-1}$ at 0 eV. The real part of $\sigma$ for SnO has two peaks at 4.09 eV and 7.27eV, where as In-doped SnO has one peak at 4.09 eV. For the case of imaginary part, σ decreases to −2.34 fs$^{-1}$ at 3.45 eV and then raises to 1.65 fs$^{-1}$ for the energy of 10.56 eV for SnO. The imaginary part of $\sigma$ is observed to vary between −0.9 to 0.82 fs$^{-1}$ for In-doped SnO.

Absorption coefficient illustrates the amount of colour absorbed per thickness of a material. The dependence of absorption coefficient with photon energy between 0 and 20 eV and with wavelength between 0 to 1000 nm is depicted in Figs. 7 (a, b), respectively. At low energies, the absorption coefficient of the films attains its minima and increases continuously with the photon energy in a manner similar to the absorption edge of a semiconductor. For pure SnO maximum absorption coefficient occurs at 7.78 eV (wavelength 162 nm), but the maxima occursat 4.44 eV (274 nm) for the In-doped SnO film, which decreases significantly with the increase in photon energy.

4. **Conclusions**

The calculated lattice parameters using the density function theory (DFT) are in good agreement with the reported experimental results with respect to bond length of SnO. The deviation of lattice parameters of SnO from the experimental data is approximately 4%, whereas for In-doped SnO this deviation becomes significant. The effect of In-doping into SnO has altered the band structure of SnO and hence decrease the band gap from 2.61 eV to 2.0 eV. The contribution of In into the SnO diminishes the static value of the refractive index and real part of the dielectric function from 1.9 to 1.4 and 3.6 to 1.97, respectively. The reflectivity, loss function, imaginary part of the refractive index and real part of the conductivity of In-doped SnO reaches to zero beyond 10.90 eV of energies, which is lower than that of SnO. Therefore, indicating that the In-doped SnO will show very high transmittance for the wave of energies greater than 10.90 eV. The above properties of SnO and In-doped SnO suggest that the films can be a suitable candidate to the researchers for applying in electronic and opto-electronic devices.

## 5. Acknowledgements

The partial financial support provided by Jagannath University, Dhaka is acknowledged to conduct this research study.

## 6. Data Availability

The datasets generated and analysed during the current study are available from the corresponding author on reasonable request.

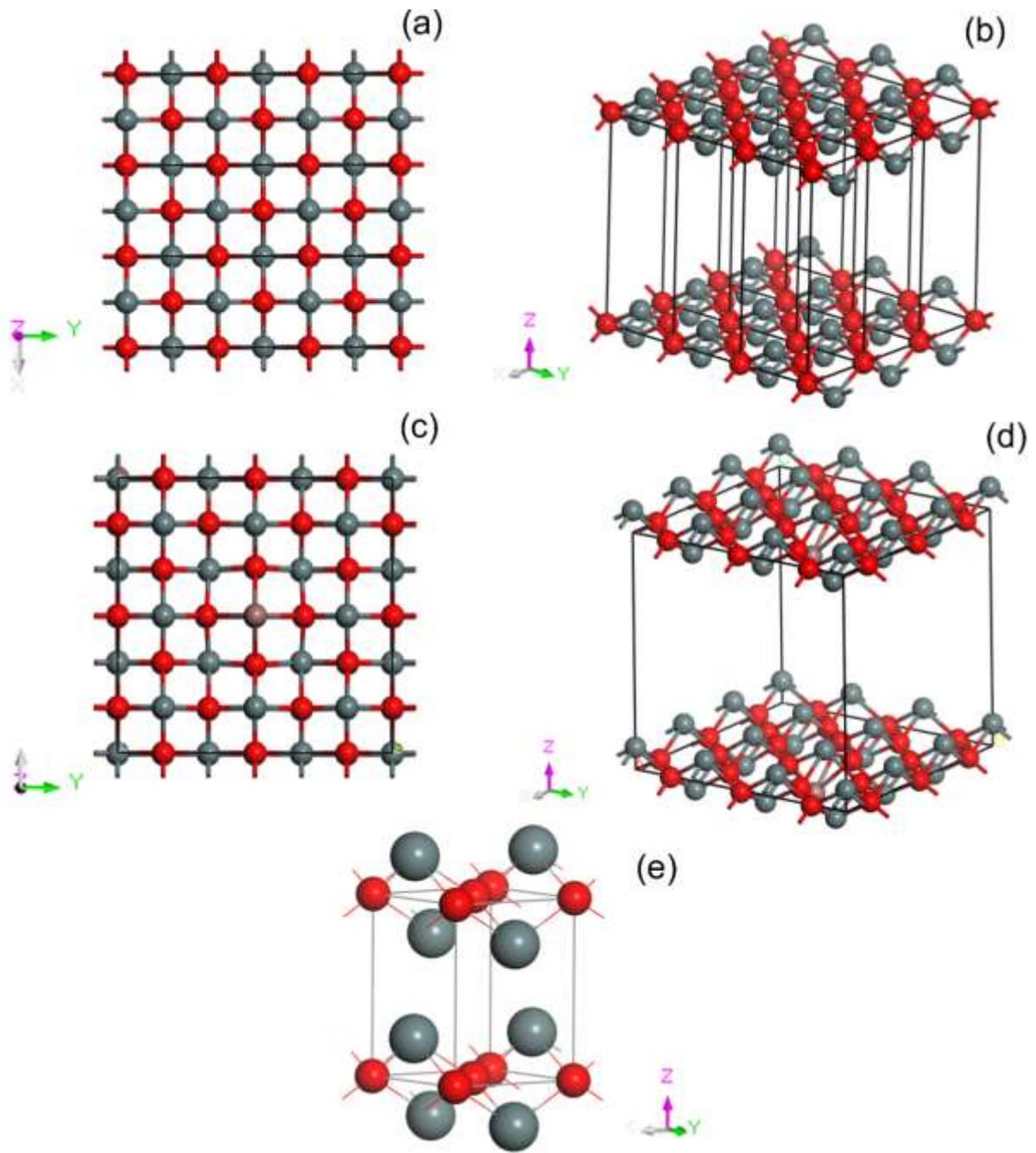

**Fig. 1** The crystal structure of SnO(2 ×2 ×1 supercell) (a) 2D and (b) 3D view, Indium doped SnO (c) 2D (d) 3D view, and (e) unit cell of SnO.

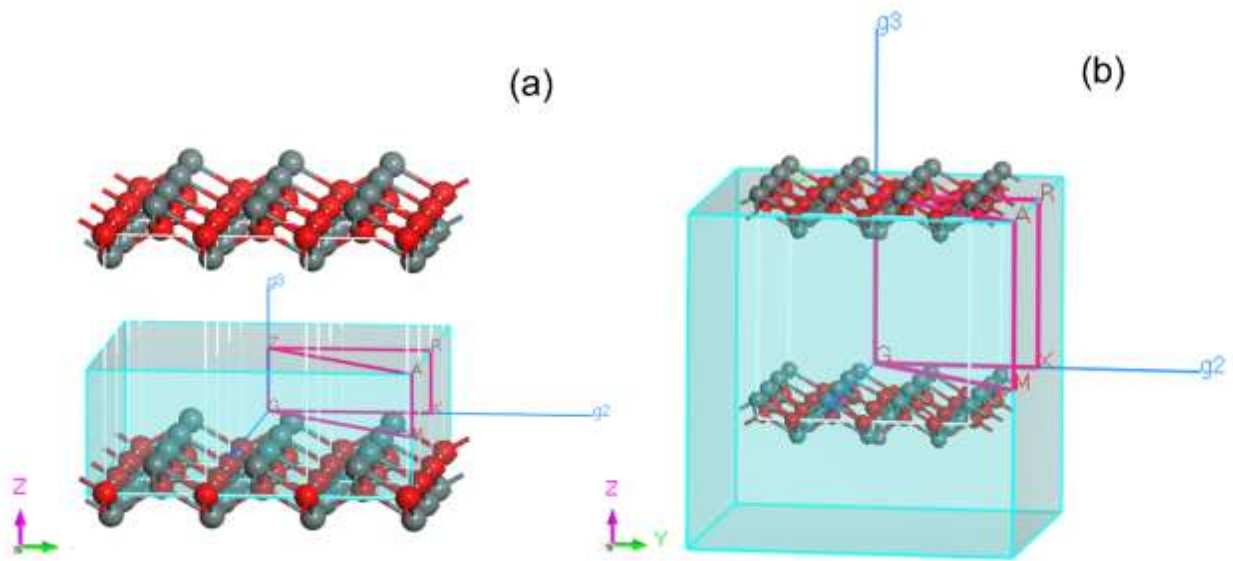

**Fig. 2** First Brillouin zone of (a) pure SnO and (b) In-doped SnO. Symmetry points are indicated to compare with the k-points shown in Fig. 3.

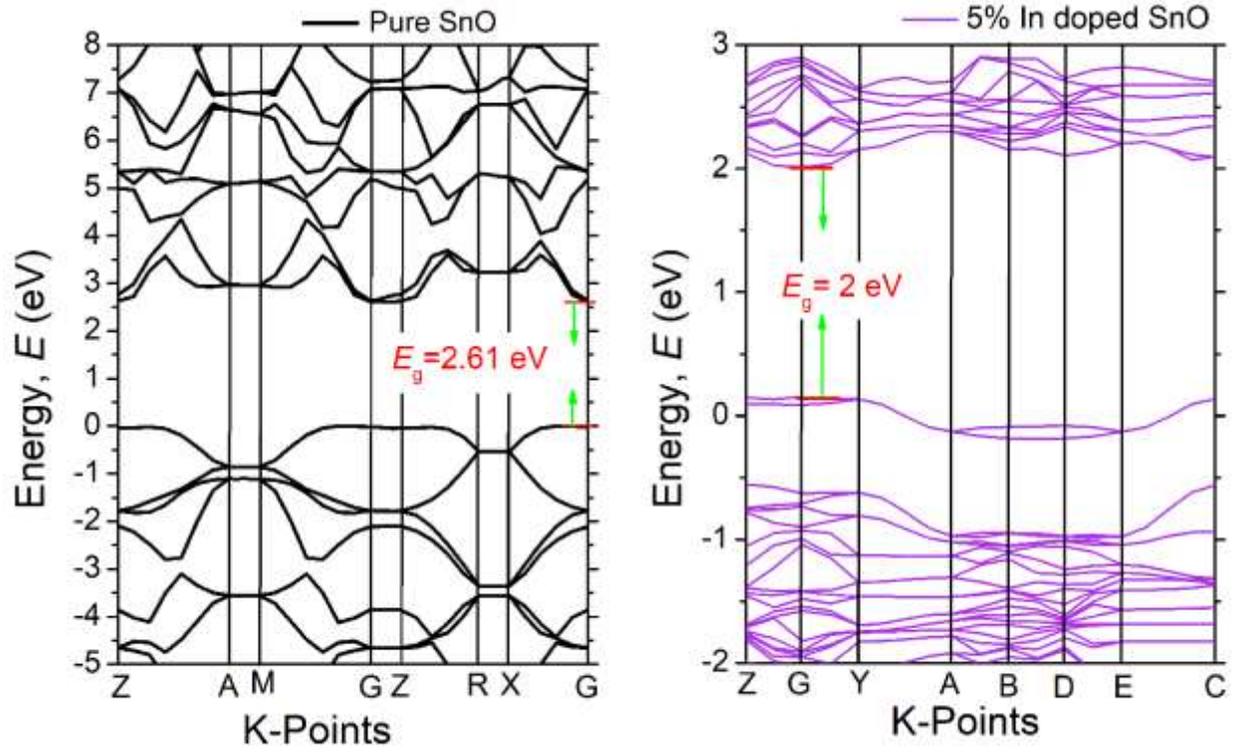

**Fig. 3** The band structures of pure SnO and In-doped SnO thin films.

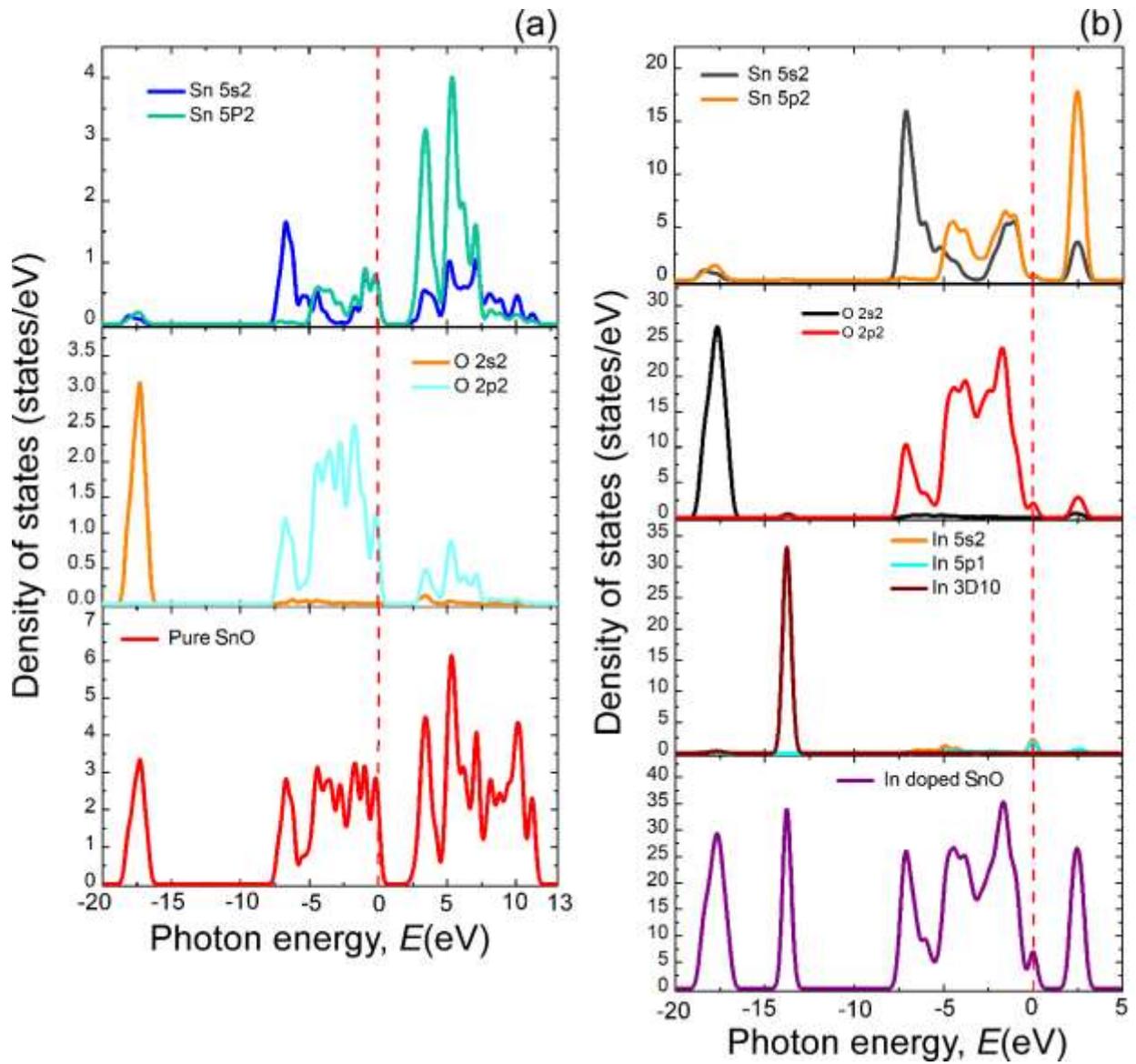

**Fig.4** The total and partial density of states of (a) SnO, (b) In-doped SnO thin films.

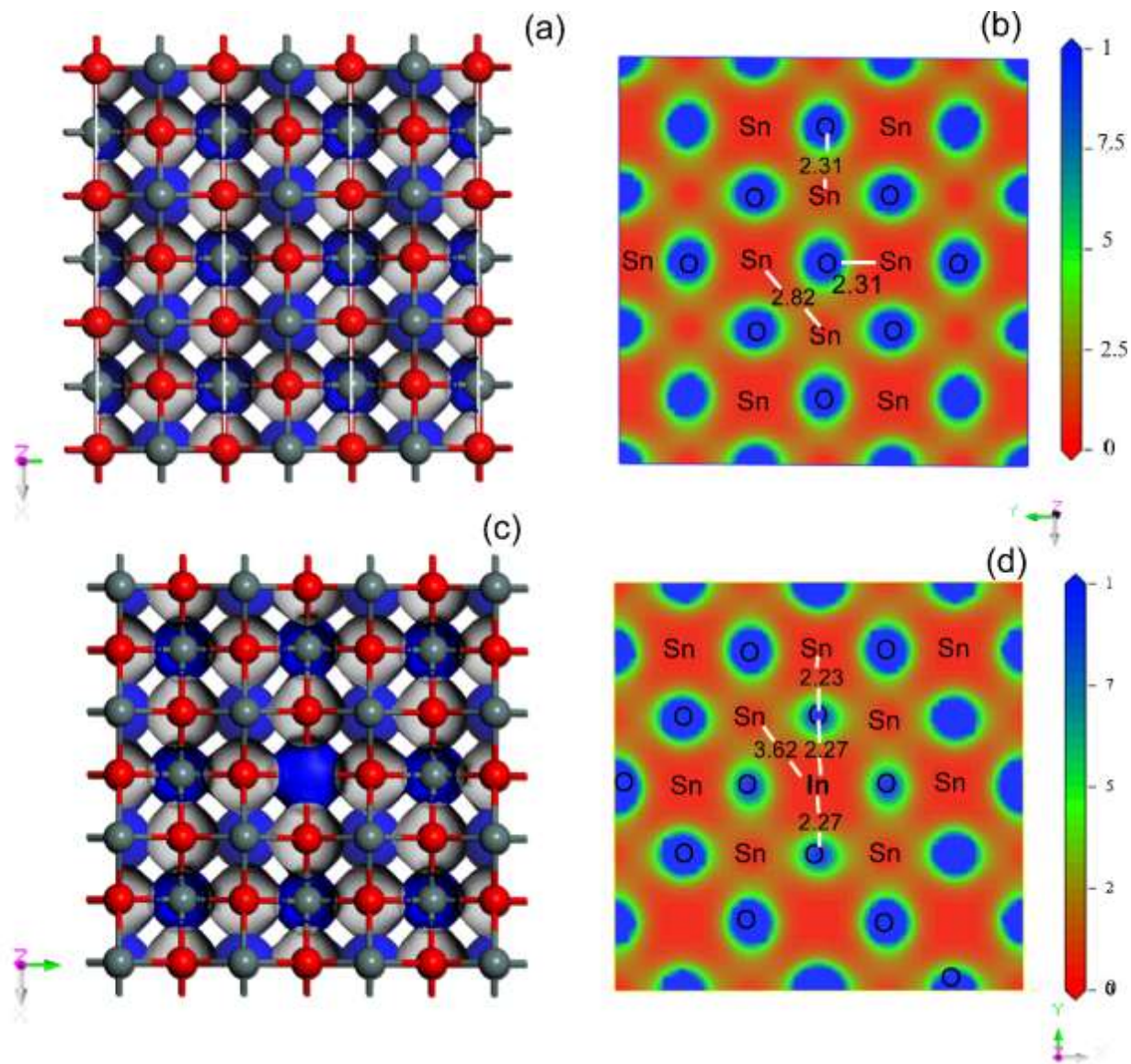

**Fig.5** Electron densities of pure SnO(a) 2D (b) color map, In-doped SnO (c) 2D (d) color map.

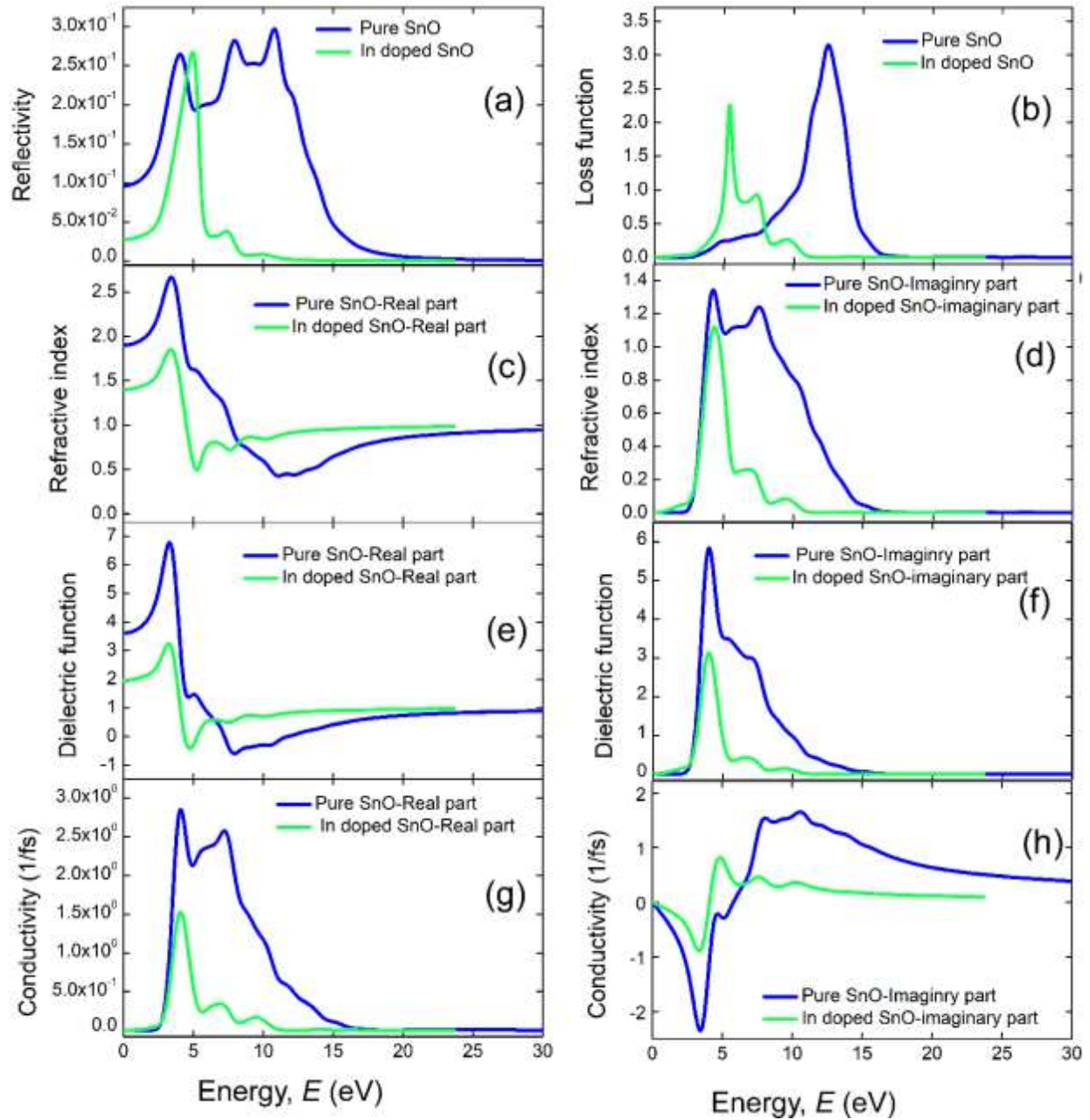

**Fig. 6** The variations of (a)reflectivity, (b)loss function, (c, e, g)real part and (d, f, h)imaginary part of the (c,d) the refractive index, (e,f)dielectric function, (g,h) conductivityof the pureSnO and In-doped SnO alongthe polarization vector [100] with photon energy .

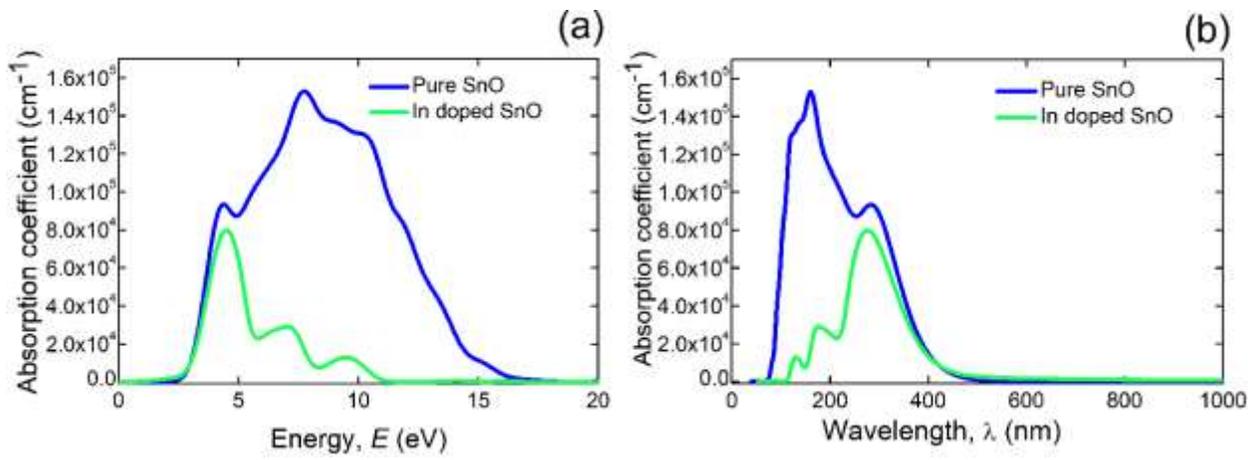

**Fig. 7** The simulated (a) photon energy dependent absorption coefficient, and (b) wavelength dependent absorption coefficient at the [100] polarization vector.

|  | Materials | *a* (Å) | *b* (Å) | *c* (Å) | Sn–O (Å) | In–O (Å) | Reference |
|---|---|---|---|---|---|---|---|
| Theoretical | SnO | 3.9651 | 3.9651 | 10.2086 | 2.285 | - | Present Work |
|  | In doped SnO | 7.2425 | 11.7754 | 12.0123 | 2.3094 | 2.266 |  |
| Experimental | SnO | 3.804 | 3.804 | 4.826 | 2.22 | - | [39] |

Table 1: Theoretical and experimental values of lattice parameters and bondlengths of SnO and In-doped SnO films.